\documentclass[twocolumn,conference]{IEEEtran}

\IEEEoverridecommandlockouts

\usepackage{bm}
\usepackage{color}
\usepackage{graphicx}
\usepackage{amsmath}
\usepackage{amssymb}
\usepackage{algorithm}
\usepackage{algorithmic}
\usepackage{ulem}
\usepackage{lineno}
\usepackage{lettrine}
\usepackage{subfigure}
\usepackage{epstopdf}
\usepackage{stfloats}
\usepackage{color}
\usepackage{graphicx}
\usepackage{amsmath}
\usepackage{amssymb}
\usepackage{algorithm}
\usepackage{algorithmic}
\usepackage{amsmath}
\usepackage{multirow}
\usepackage{booktabs}
\usepackage{array}
\usepackage{amsthm}
\usepackage{stfloats}
\usepackage{caption}
\usepackage{subfigure}
\usepackage{bm}
\usepackage{booktabs}
\usepackage{setspace}
\usepackage{makecell}

\newcommand{\be}{\begin{equation}}
	\newcommand{\ee}{\end{equation}}
\newcommand{\bea}{\begin{eqnarray}}
	\newcommand{\eea}{\end{eqnarray}}
\newcommand{\ba}{\begin{array}}
	\newcommand{\ea}{\end{array}}

\newcommand{\non}{\nonumber}

\title{BS-RIS-User Association and Beamforming Designs for RIS-aided Cellular Networks}
\author{\IEEEauthorblockN{Sifan Liu$^{\dag}$, Pengfei Ni$^{\dag}$, Rang Liu$^{\dag}$, Yang Liu$^{\dag }$, Ming Li$^{\dag \ast}$, and Qian Liu$^{\ddag}$
		\vspace{-0.0 cm} }
	
	\IEEEauthorblockA{$^{\dag}$ School of Information and Communication Engineering   \\  Dalian University of Technology, Dalian, Liaoning 116024, China \\ E-mail: \texttt{\{sifanliu,pfni,liurang\}@mail.dlut.edu.cn, \{yangliu\_613,mli\}@dlut.edu.cn} }
	\IEEEauthorblockA{$^{\ddag}$ School of Computer Science and Technology \\  Dalian University of Technology, Dalian, Liaoning 116024, China \\ E-mail: \texttt{qianliu@dlut.edu.cn}   }}
\begin{document}
	\maketitle
	\pagestyle{empty}
	\thispagestyle{empty}
	\begin{abstract}
		Reconfigurable intelligent surface (RIS) has been regarded as a revolutionary and promising technology owing to its powerful feature of adaptively shaping wireless propagation environment.
		However, as a frequency-selective device, the RIS can only effectively provide tunable phase-shifts for signals within a certain frequency band.
		Thus, base-station (BS)-RIS-user association is an important issue to maximize the efficiency and ability of the RIS in cellular networks.
		In this paper, we consider a RIS-aided cellular network and aim to maximize the sum-rate of downlink transmissions by designing BS-RIS-user association as well as the active and passive beamforming of BSs and RIS, respectively.
		A dynamically successive access algorithm is developed to design the user association.
		During the dynamical access process, an iterative algorithm is proposed to alternatively obtain the active and passive beamforming.
		Finally, the optimal BS-RIS association is obtained by an exhaustive search method.
		Simulation results illustrate the significant performance improvement of the proposed BS-RIS-user association and beamforming design algorithm.
	\end{abstract}	
	
	
	\section{Introduction}
	To satisfy the increasing requirements for higher data rate and capacity of the fifth generation (5G) wireless communication systems, several key enabling technologies, e.g., massive multi-input multi-output (MIMO) and ultra-dense network, are employed in recent years.
	However, the deployment of numerous base-stations (BSs), relays, and large-scale antenna arrays incurs increasing energy consumption and hardware cost.
	Recently, reconfigurable intelligent surface (RIS) has been introduced as a revolutionary and promising technology for future beyond 5G and 6G networks \cite{M. D. Renzo}, \cite{Q. Wu overview}.
	
	RIS is a planar surface that consists of massive passive elements, which are composed of the configurable electromagnetic (EM) internals.
	The amplitude and phase shift of the incident EM waves can be adjusted to enable the reconfiguration of the wireless environment, which provides new degrees of freedom (DoFs) for optimizing wireless networks \cite{X. Yu}.
	Owing to the passive elements and the simple hardware structure, RIS can provide significant performance improvement with low power consumption and hardware cost, which widely attracts the research interest from both academia and industry.
	
	Plenty of works \cite{X. Yu}-\cite{C. Huang} have been proposed to optimize RIS-assisted communication systems, which usually has only one transmitter/BS.
	For more complicated multi-BS cellular networks, the user association is one of the important procedures to further enhance the spectral efficiency \cite{A. Alizadeh}.
	Thus, the authors in \cite{W. Mei1} and \cite{D. Zhao} considered the optimization of RIS beamforming in a multi-BS cellular system and investigated the RIS-user and BS-user association, respectively.
	Furthermore, relationship between BSs and RISs are studied in \cite{W. Mei2}, which introduces a BS-RIS-user association method to maximize the utility of users in a single-input single-output (SISO) downlink wireless network.
	However, above-mentioned works ignore an important fact that the RIS is a frequency-selective device \cite{W. Cai}, \cite{H. Li}.
	In other words, RIS can only provide tunable phase-shifts for signals within a certain frequency band while generates the fixed phase-shifts for the signals of other frequency bands.
	Therefore, when multiple BSs operate at different frequency bands, RIS can only effectively serve one BS and construct favorable channels between this BS and its associated users. Consequently, the BS-RIS association is indeed a crucial problem worth studying.
	
	In this paper, we consider the RIS-aided cellular network and aim to maximize the sum-rate by jointly optimizing the BS-RIS-user association, the active beamforming at BSs, and the passive beamforming of RIS.
	To tackle this difficult non-convex problem, we first assume that each BS is served by an individual RIS and propose a dynamically successive access algorithm to obtain the BS-user association.
	During the dynamical access process, an iterative algorithm is utilized to alternatively design the active beamforming of BS and the passive beamforming of RIS.
	To be specific, we first employ the Lagrangian dual transform method to transform the original problem into a multi-ratio fractional programming (FP) problem.
	After that, we design the active beamforming by the normalized zero-forcing (ZF) precoding technology and optimize the passive beamforming by an iterative reflection coefficient updating method.
	Finally, the optimal BS-RIS association can be obtained by searching from all the possibilities. Simulation results illustrate that the proposed algorithm has the significant performance improvement compared to the comparison algorithms.
	
	\section{System Model and Problem Formulation}
	\subsection{System Model}
\begin{figure}[!t]
		\centering
		\includegraphics[width =2.65 in]{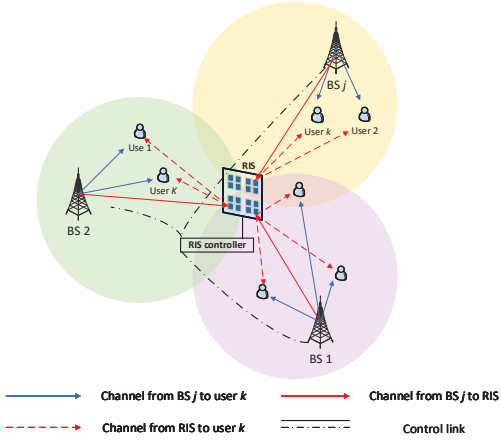}
		\vspace{-0.15 cm}
		\caption{RIS-assisted cellular system.}
		\vspace{-0.75 cm}
		\label{fig:system}
	\end{figure}

	We consider a RIS-aided cellular system with one RIS, a set of single-antenna users $\mathcal{K} = \{1,2,...,K\}$, and a set of BSs $\mathcal{J} = \{1,2,...,J\}$, as shown in Fig. \ref{fig:system}.
	Each BS is equipped with $M$ antennas and can simultaneously transmit independent data streams to a maximum of $K$ users, $M \geq K$.
	The RIS consists of a set of reflection elements $\mathcal{N} = \{1,2,...,N\}$, which are controlled by the BSs with wireless or wired links.
	We assume that each BS operates at different frequency bands to serve its associated users.
	
	Since a RIS is deployed in the cellular system, each BS-user link is connected not only by the direct BS-user channel, but also via the cascaded BS-RIS-user channel which depends on the passive beamforming of the RIS.
	Therefore, it is crutial to investigate the joint design of user association, BS active beamforming, and RIS passive beamforming.
	To explicitly express the association between the users and the BSs, we define the user association matrix $\mathbf{U}$,  which consists of binary variables $u_{j,k}$:
	\vspace{-0.15 cm}
	\begin{equation}
		\begin{small}
			u_{j, k}=\left\{\begin{array}{l}
				\hspace{-0.1 cm} 	1, \text { user-} k \text { is served by BS-} j,\forall j \in \mathcal{J}, \forall k \in \mathcal{K}. \\
				\hspace{-0.1 cm} 	0, \text { otherwise}.
			\end{array}\right.
		\end{small}
	\end{equation}
	The $j$-th row and the $k$-th column of $\mathbf{U}$ represent the user association vector at BS-$j$ and the BS association vector for user-$k$, respectively.
	
	As a frequency-selective device, the RIS can only optimize its passive beamforming for the associated BS while reflecting the signals from other BSs with fixed phase-shifts \cite{W. Cai}, \cite{H. Li}.
	This fact means that the RIS can only effectively serve one BS and its associated users by constructing tunable channels between them.
	Thus, in order to illustrate the association between BSs and the RIS, we define a RIS association vector $\mathbf{r}=[r_{1},...,r_{J}]^{T}$.
	Moreover, based on the practical RIS model, the phase-shift matrix can be written as $\mathbf{\Phi}_{j}\triangleq\operatorname{diag}(\boldsymbol{\varphi}_{j}),\forall j \in \mathcal{J}$, in which $\boldsymbol{\varphi}_{j}=[\varphi_{j,1} ,..., \varphi_{j,N}]^{T}=[\beta_{1} e^{\jmath \theta_{1} r_{j}},..., \beta_{N} e^{\jmath \theta_{N} r_{j}}]^{T} \in \mathbb{C}^{N \times 1}$, $\beta_{n}=1$ and $\theta_{n} \in [0,2\pi)$ are the control variables of the amplitude and phase of the RIS reflected signal, respectively.
	In particular, if the RIS assists BS-$j$, $r_{j} = 1$, the phase-shifts in $\mathbf{\Phi}_{j}$ is tunable and can be optimized; otherwise, $r_{j} = 0$, the phase-shift $\theta_{n}$ is fixed as $2\pi$ and $\varphi_{j,n} = 1,\forall n \in \mathcal{N}$. Moreover, the constraint $\sum_{j \in \mathcal{J}} r_{j}=1$ should be satisfied because of the frequency selectivity of the RIS.
	
	Then, we denote $s_{j,k}$ as the transmitted symbols for user-$k$ served by BS-$j$, which satisfies $\mathbb{E}\{s_{j,k}s_{j,k}^*\}=1$ and $\mathbb{E}\{s_{j,k}s_{j,i}^*\}=0$, $\forall i\neq k$.
	Let $\mathbf{w}_{j,k}$ be the beamforming vector for BS-$j$ to user-$k$.
	Then, the received signal of user-$k$ can be expressed as
	\vspace{-0.15 cm}
	\begin{equation}
		\begin{small}
			\label{y_k}
			\begin{aligned}
				y_{k}=&\underbrace{\sum_{j \in \mathcal{J}}u_{j, k} \left(\mathbf{h}_{\text{d}, j, k}^{H}+ \mathbf{h}_{\text{r}, k}^{H} \mathbf{\Phi}_{j} \mathbf{G}_{j}\right) \mathbf{w}_{j, k} s_{j, k}}_{\text {Useful signals for user-\textit{k}}}\\
				&+\underbrace{\sum_{j \in \mathcal{J}} u_{j, k}\hspace{-0.3cm}\sum_{i \in \mathcal{A}_{j} ,i \neq k} \hspace{-0.3cm}\left(\mathbf{h}_{\text{d}, j, k}^{H}+\mathbf{h}_{\text{r}, k}^{H} \mathbf{\Phi}_{j} \mathbf{G}_{j}\right) \mathbf{w}_{j, i} s_{j, i}}_{\text {Interference from other users }}+n_{k}, k \in \mathcal{K},
			\end{aligned}
		\end{small}
	\end{equation}
	where $\mathbf{h}_{\text{d},j,k}\in \mathbb{C}^{M \times 1}$, $\mathbf{G}_{j} \in \mathbb{C}^{N \times M}$, $\mathbf{h}_{\text{r},k} \in \mathbb{C}^{N \times 1}$, $j \in \mathcal{J}, k \in \mathcal{K}$ are direct channels from BS-$j$ to user-$k$, from BS-$j$ to RIS, and from RIS to user-$k$, respectively.
	$\mathcal{A}_{j} \subset \mathcal{K}$ denotes the set of users which are associated with BS-$j$.
	$n_{k} \sim \mathcal{CN} (0, \sigma_{k}^{2})$ is the complex additive white Gaussian noise with variance $\sigma_{k}^{2}$ and zero mean.
	Besides, we assume that the channel state information (CSI) of all the channels are prefectly known by BSs.
	
	According to above descriptions, the signal-to-interference-plus-noise ratio (SINR) of user-$k$ can be written as
	\vspace{-0.15 cm}
	\begin{equation}
		\label{sinr_k}
		\begin{small}
			\begin{aligned}
				\gamma_{k}\hspace{-0.1 cm}=&\sum\limits_{j \in \mathcal{J}} u_{j, k} \frac{|\widetilde{\mathbf{h}}_{j,k}^{H} \mathbf{w}_{j, k}|^{2}}{\hspace{-0.2 cm}\sum\limits_{i \in \mathcal{A}_{j}, i \neq k}\hspace{-0.1 cm} |\widetilde{\mathbf{h}}_{j,k}^{H} \mathbf{w}_{j, i}|^{2}+\sigma_{k}^{2}}, k \in \mathcal{K},
			\end{aligned}
		\end{small}
	\end{equation}
	where $\widetilde{\mathbf{h}}_{j,k}^{H}\triangleq\mathbf{h}_{\text{d}, j, k}^{H}+\mathbf{h}_{\text{r}, k}^{H} \mathbf{\Phi}_{j} \mathbf{G}_{j}$ denotes the equivalent channel from BS-$j$ to user-$k$ for brevity. Hence, the achievable rate of user-$k$ can be calculated by
	\vspace{-0.15 cm}
	\begin{equation}\label{r_k}
		R_{k} = \log_{2} \left(1+\gamma_{k}\right),k \in \mathcal{K}.
	\end{equation}
	
	\subsection{Problem Formulation}
	\begin{spacing}{1.0}
		We aim to maximize the achievable sum-rate of all users by jointly optimizing the user association matrix $\mathbf{U}$, the RIS association vector $\mathbf{r}$, the active beamforming $\mathbf{w}_{j,k}$ at the BSs, and the passive beamforming $\mathbf{\Phi}_{j}$ of RIS, subject to the constraints of the transmit power at BSs and phase-shifts at RIS.
		Thus, the sum-rate maximization problem is formulated as\end{spacing}
	\vspace{-0.5cm}
	\begin{subequations}\label{P1}
		\begin{small}
			\begin{align}
				\mathrm{P} 1: \quad \mathop{\max}\limits_{\substack{\mathbf{U},\mathbf{r},\mathbf{w}_{j,k},\mathbf{\Phi}_{j}\\ \forall j \in \mathcal{J}, \forall k \in \mathcal{K}}} \;
				&f_{1}(\mathbf{U},  \mathbf{r},\mathbf{w}_{j,k},\mathbf{\Phi}_{j})\hspace{-0.05 cm}=\hspace{-0.05 cm}  \sum_{k \in \mathcal{K}}\hspace{-0.1 cm} R_{k},\\
				\text {s.t.} \quad
				& \sum_{j \in \mathcal{J}}\sum_{k \in \mathcal{K}} u_{j,k}\left\|\mathbf{w}_{j,k}\right\|^{2} \le P_{\text{max}}, \label{Pt}\\
				& \theta_{n} \in \left[0,2\pi\right), \quad \forall n \in \mathcal{N}, \label{fain}\\	
				& u_{j,k} = \{0,1\},  \quad \forall j \in \mathcal{J},\forall k \in \mathcal{K},\label{uone}\\
				& \sum_{j \in \mathcal{J}} u_{j, k}=1, \quad \forall k \in \mathcal{K} \label{oneuser},\\
				& \sum_{k \in \mathcal{K}} u_{j, k} \geqslant 1, \quad \forall j \in \mathcal{J} \label{BS},\\
				& r_{j}  = \{0,1\}, \quad \forall j \in \mathcal{J},\label{rone}\\
				& \sum_{j \in \mathcal{J}} r_{j}=1,  \label{RIS}
			\end{align}
		\end{small}
	\end{subequations}\hspace{-0.14cm}where $P_{\text{max}}$ is the maximum transmit power for all BSs.
	
	Due to the non-convex function $f_{1}(\mathbf{U},\mathbf{r},\mathbf{w}_{j,k},\mathbf{\Phi}_{j})$ and the discrete values of binary variables, i.e., $u_{j,k}$ and $r_{j}$, it is intractable to solve problem ($\mathrm{P} 1$) directly.
	Thus, in the following part, we decompose ($\mathrm{P} 1$) into sub-problems and propose a dynamically successive access algorithm, an iterative beamforming design algorithm, and an exhaustive search method to solve the sum-rate maximization problem ($\mathrm{P} 1$).
	
	\section{Association and Beamforming Design}
	In this section, we introduce the proposed dynamically successive access algorithm, the alternatively iterative algorithm, and the exhaustive search method to solve the original problem ($\mathrm{P} 1$).
	To be specific, we decompose problem ($\mathrm{P} 1$) into four sub-problems, i.e., BS active beamforming, RIS passive beamforming, user association, and RIS association optimization.
	\subsection{BS Active Beamforming Design}
	In this subsection, we present the BS active beamforming design.
	With $\mathbf{U}$, $\mathbf{r}$, and $\mathbf{\Phi}_{j}$ being fixed, the original problem can be reformulated as
	\begin{subequations}\label{PW}
		\begin{align}
			\mathrm{P} 2:  \mathop{\max}\limits_{\substack{\mathbf{w}_{j,k}\\ \forall j \in \mathcal{J}, \forall k \in \mathcal{K} }} \; &f_{2}(\mathbf{w}_{j,k})=  \sum_{k \in \mathcal{K}} R_{k},\\
			\text {s.t.} \quad ~
			& \sum_{j \in \mathcal{J}}\sum_{k \in \mathcal{K}} u_{j,k}\left\|\mathbf{w}_{j,k}\right\|^{2} \le P_{\text{max}},\label{Pmax}
		\end{align}
	\end{subequations}
	where the set of associated users $\mathcal{A}_j$ is known here by the fixed $\mathbf{U}$.
	
	The main purpose of designing the active beamformer $\mathbf{w}_{j,k}$ is to suppress the intra-cell interference.
	Fortunately, previous works have shown that using suitable precoding strategies can effectively eliminate intra-cell interference \cite{E}. Inspired by this, we use the normalized zero-forcing (ZF) precoding to cancel the interference.
	To be specific, we define $\mathbf{H}_{j} \triangleq [\widetilde{\mathbf{h}}_{j,1},...,\widetilde{\mathbf{h}}_{j,k},...,\widetilde{\mathbf{h}}_{j,|\mathcal{A}_j|}],\forall j \in \mathcal{J},\forall k \in \mathcal{A}_{j}$ as the channel from BS-$j$ to the set of users $\mathcal{A}_{j}$ and
	$\mathbf{f}_{j,k}$ as the $k$-th row of ${\mathbf{H}_{j}}^{\dag}$ for brevity. Then, the beamforming vector $\mathbf{w}_{j,k}$ can be expressed as
	\begin{equation}\label{w}
		\begin{aligned}
			\mathbf{w}_{j, k}&=\sqrt{p_{j,k}} \frac{\mathbf{f}_{j,k}^H}{\big\|\mathbf{f}_{j,k}^H\big\|},\forall k \in \mathcal{A}_{j},\forall j \in \mathcal{J},
		\end{aligned}
	\end{equation}
	where $p_{j,k} = P_{\text{max}}/{K}$ is the available power for the transmission signal from BS-$j$ to user-$k$, which satisfies the constraint \eqref{Pmax}.
	
	\subsection{RIS Passive Beamforming Optimization}
	In this subsection, we introduce the iterative reflection coefficient updating method to optimize the passive beamforming.
	Notice that, during the procedure of user association, we dynamically update the active beamforming $\mathbf{w}_{j,k}, \forall k \in \mathcal{A}_{j}$ and passive beamforming ${\mathbf{\Phi}}_{j}$.

	Since the optimization of ${\mathbf{\Phi}}_{j}$ will only affect the transmission rate of the associated users, problem ($\mathrm{P} 1$) can be rewritten as
	\vspace{-0.05cm}
	\begin{subequations}\label{P3}
		\begin{align}
			\mathrm{P} 3:\quad \max \limits_{{\mathbf{\Phi}}_{j}} \quad & f_{3}(\mathbf{\Phi}_{j})= \sum_{l \in \mathcal{A}_{j}} R_{l}, \\
			\text { s.t. } \quad &\theta_{n} \in \left[0,2\pi\right), \quad \forall n \in \mathcal{N},
		\end{align}
	\end{subequations}
	where the SINR of the related users is
	\begin{equation}\label{sinr_l}
		\gamma_{l}=\frac{\left|\widetilde{\mathbf{h}}_{j, l}^{H} \mathbf{w}_{j, l}\right|^{2}}{\sum_{i \in \mathcal{A}_{j},i \neq l}\left| \widetilde{\mathbf{h}}_{j, l}^{H} \mathbf{w}_{j, i}\right|^{2}+\sigma_{l}^{2}}, l \in \mathcal{A}_{j},
	\end{equation}
	and $\widetilde{\mathbf{h}}_{j, l}$ represents the equivalent channel from BS-$j$ assisted by RIS to user-$l$.
	
	Due to the logarithm term, problem ($\mathrm{P} 3$) is still a non-convex problem, which is intractable to solve.
	Inspired by \cite{K. Shen}, we use the Lagrangian dual transform method to move $\gamma_{l}$ outside the logarithm in $f_{3}(\mathbf{\Phi}_{j})$ to obtain a more tractable form of the objective function, which is given by
	\begin{equation}\label{f3'}
		\begin{aligned}
		\hspace{-0.1cm}
			f_{3'}(\mathbf{\Phi}_{j}, \boldsymbol{\lambda})= &\sum_{l \in \mathcal{A}_{j}}  \log _{2}\left(1+\lambda_{l}\right) \\
			&- \frac{1}{\ln2}\bigg(\sum_{l \in \mathcal{A}_{j}}  \lambda_{l}- \sum_{l \in \mathcal{A}_{j}}  \frac{\left(1+\lambda_{l}\right) \gamma_{l}}{1+\gamma_{l}}\bigg),
		\end{aligned}
	\end{equation}
	where $\boldsymbol{\lambda} \triangleq [\lambda_{1},...,\lambda_{l},...,\lambda_{|\mathcal{A}_j|}]$ is an auxiliary vector.
	When $\mathbf{\Phi}_{j}$ is fixed, it is obvious that $f_{3'}(\mathbf{\Phi}_{j}, \boldsymbol{\lambda})$ is a concave function of $\boldsymbol{\lambda}$, so we can easily obtain the optimal solution of $\lambda_{l}$ by setting $\partial f_{3'} / \partial \lambda_{l}=0,\forall l \in  \mathcal{A}_{j}$ as
	\begin{equation}\label{lamda}
		\lambda_{l}^{\text{opt}} = \gamma_{l}, \quad l \in \mathcal{A}_{j}.
	\end{equation}
	With fixed $\boldsymbol{\lambda}$ and removing irrelevant terms, problem ($\mathrm{P} 3$) can be transformed into
	\begin{subequations}\label{P4}
		\begin{align}
			\mathrm{P} 4: \quad \max\limits_{\mathbf{\Phi}_{j}} \quad &f_{4}(\mathbf{\Phi}_{j})=\sum_{l \in \mathcal{A}_{j}} \frac{\left(1+\lambda_{l}\right) \gamma_{l}}{1+\gamma_{l}},\label{f4}\\
			\text { s.t. } \quad &\theta_{n} \in \left[0,2\pi\right), \quad \forall n \in \mathcal{N}.
		\end{align}
	\end{subequations}
	Since $f_{4}(\mathbf{\Phi}_{j})$ is the sum of multiple fractional equations, it is obvious that ($\mathrm{P} 4$) is a multi-ratio FP problem.
	To tackle problem ($\mathrm{P} 4$), we firstly substitute \eqref{sinr_l} into \eqref{f4} and introduce an auxiliary variable $\mathbf{q} \triangleq [q_{1},...,q_l,...,q_{|\mathcal{A}_j|}]$ to decouple the numerators and denominators based on the quadratic transform proposed in \cite{K. Shen}.
	Moreover, by defining $\mathbf{a}_{i, l}\triangleq\operatorname{diag}(\mathbf{h}_{\text{r}, l}^{H}) \mathbf{G}_{j} \mathbf{w}_{j,i}$, $b_{i, l}\triangleq\mathbf{h}_{\text{d}, j, l}^{H} \mathbf{w}_{j,i}$, ($\mathrm{P} 4$) can be rewritten as
	\begin{subequations}\label{P4'}
		\begin{align}
			\mathrm{P} 4': \quad \max\limits_{\boldsymbol{\varphi}_{j}} \quad &
			f_{4'}(\boldsymbol{\varphi}_{j}, \mathbf{q}),\\
			\text { s.t. } \quad &\theta_{n} \in \left[0,2\pi\right), \quad \forall n \in \mathcal{N},
		\end{align}
	\end{subequations}
	where the transformed objective function is
	\begin{equation}\label{f4'}
			\begin{aligned}
				f_{4'}(\boldsymbol{\varphi}_{j}, \mathbf{q})=\hspace{-0.1 cm}  &\sum_{l \in \mathcal{A}_{j}} 2 \sqrt{1+\lambda_{l}} \mathfrak{Re} \left\{q_{l}^{*} \boldsymbol{\varphi}_{j}^{H} \mathbf{a}_{l, l}+q_{l}^{*} b_{l, l}\right\} \\
				&-\hspace{-0.1 cm} \sum_{l \in \mathcal{A}_{j}}\hspace{-0.1 cm} \left|q_{l}\right|^{2}\hspace{-0.1 cm}\bigg(\sum_{i \in \mathcal{A}_{j}}\hspace{-0.1 cm} \left|b_{i, l}+\boldsymbol{\varphi}_{j}^{H} \mathbf{a}_{i, l}\right|^{2}\hspace{-0.1 cm} + \sigma_{l}^{2}\bigg).\\
			\end{aligned}
	\end{equation}	
	Obviously, the objective function \eqref{f4'} is a convex problem with respect to $\boldsymbol{\varphi}_{j}$ and $\mathbf{q}$ separately.
	Then, we can solve problem ($\mathrm{P} 4'$) by optimizing $\boldsymbol{\varphi}_{j}$ and $\mathbf{q}$ alternatively.
	For given $\boldsymbol{\varphi}_{j}$, the optimal $\mathbf{q}$ can be found by setting $\partial f_{4'} / \partial q_{l}$ to zero, and the solution is
	\begin{equation}\label{q}
		\begin{small}
			\begin{aligned}
				q_{l}^{\text{opt}}&=\frac{\sqrt{1+\lambda_{l}}\left(b_{l, l}+\boldsymbol{\varphi}_{j}^{H} \mathbf{a}_{l, l}\right)}{\sum\limits_{i \in \mathcal{A}_{j}}\left(\boldsymbol{\varphi}_{j}^{H} \mathbf{a}_{i, l} \mathbf{a}_{i, l}^{H} \boldsymbol{\varphi}_{j}+2 \operatorname{Re}\left\{b_{i, l}^{*} \boldsymbol{\varphi}_{j}^{H} \mathbf{a}_{i, l}\right\}+\left|b_{i, l}\right|^{2}\right)+\sigma_{l}^{2}}.
			\end{aligned}
		\end{small}
	\end{equation}
	Then, for given $\mathbf{q}$, after removing irrelevant constant terms, problem ($\mathrm{P} 4'$) can be rewritten as
	\begin{subequations}\label{P5}
		\begin{align}
			\mathrm{P} 5: \quad \max\limits_{\boldsymbol{\varphi}_{j}} \quad &
			f_{5}(\boldsymbol{\varphi}_{j}) \hspace{-0.1cm}\triangleq -\boldsymbol{\varphi}_{j}^{H} \mathbf{D} \boldsymbol{\varphi}_{j}+2 \mathfrak{Re} \left\{\boldsymbol{\varphi}_{j}^{H} \mathbf{v}\right\},\\
			\text { s.t. } \quad &\theta_{n} \in \left[0,2\pi\right), \quad \forall n \in \mathcal{N},
		\end{align}
	\end{subequations}
	\hspace{-0.15cm}in which we define
	\begin{subequations}
		\begin{align}
			\mathbf{D} &\triangleq\sum_{l \in \mathcal{A}_{j}}\left|q_{l}\right|^{2} \sum_{i \in \mathcal{A}_{j}} \mathbf{a}_{i, l} \mathbf{a}_{i, l}^{H},\\ \label{u}		
			\mathbf{v} &\triangleq\sum_{l \in \mathcal{A}_{j}}\bigg(\sqrt{1+\lambda_{l}} q_{l}^{*} \mathbf{a}_{l, l}-\left|q_{l}\right|^{2} \sum_{i \in \mathcal{A}_{j}} b_{i, l}^{*} \mathbf{a}_{i, l}\bigg),
		\end{align}
	\end{subequations}
	for brevity. 
	
	This problem can be directly solved by the Riemannian conjugate gradient-based algorithm.
	However, in order to reduce the computational complexity, we use the iterative reflection coefficient updating method to iteratively solve each element of $\boldsymbol{\varphi}_{j}$ until convergence \cite{H. Guo}.
	Specifically, we first denote $d_{i,m}$ as the element at the $i$-th row and the $m$-th column of $\mathbf{D}$, and $v_{i}$ as the $i$-th element of $\mathbf{v}$. Since $\mathbf{D}$ is a Hermitian matrix, i.e., $d_{i, m}=d_{m, i}^{*}$, we can obtain
	\begin{equation}\label{faiv}
			\begin{aligned}
				\hspace{-2.4cm}\boldsymbol{\varphi}_{j}^{H} \mathbf{v}=\varphi_{j,n}^{*} v_{n}+\sum_{i \in \mathcal{N}, i \neq n} \varphi_{j,i}^{*} v_{i},
			\end{aligned}
	\end{equation}
	\begin{equation}
			\begin{aligned}
				\boldsymbol{\varphi}_{j}^{H} \mathbf{D} \boldsymbol{\varphi}_{j}= \varphi_{j,n}^{*} d_{n, n} \varphi_{j,n} +2 \mathfrak{Re}\bigg\{\hspace{-0.05 cm} \sum_{m \in \mathcal{N}, m \neq n}\hspace{-0.4 cm} \varphi_{j,n}^{*} d_{n, m} \varphi_{j,m}\hspace{-0.05 cm} \bigg\} \\ \non
			\end{aligned}
	\end{equation}
	\vspace{-0.6cm}
	\begin{equation}\label{faivfai}
			\begin{aligned}
				\hspace{-0.1 cm} + \sum_{i \in \mathcal{N}, i \neq n} \sum_{m \in \mathcal{N}, m \neq n} \hspace{-0.3 cm}\varphi_{j,i}^{*} d_{i, m} \varphi_{j,m}.
			\end{aligned}
	\end{equation}
	By substituting \eqref{faiv} and \eqref{faivfai} into \eqref{P5} and removing irrelevant constants, the objective function $f_{5}(\boldsymbol{\varphi}_{j})$ can be further transformed into
	\vspace{-0.1cm}
	\begin{equation}\label{f5'}
		\begin{small}
			\begin{aligned}
				\hspace{-0.45 cm}f_{5'}\left(\varphi_{j,n}\right)&=-\hspace{-0.1 cm}\left|\varphi_{j,n}\right|^{2} \hspace{-0.05 cm}d_{n, n}\hspace{-0.05 cm}+\hspace{-0.05 cm}2 \mathfrak{Re}\hspace{-0.05 cm}\bigg\{\hspace{-0.05 cm}\varphi_{j,n}^{*}\hspace{-0.1 cm}\bigg(v_{n}\hspace{-0.1 cm}-\hspace{-0.5 cm}\sum_{m \in \mathcal{N}, m \neq n}\hspace{-0.45 cm} d_{n, m} \varphi_{j,m}\hspace{-0.1 cm}\bigg)\hspace{-0.1 cm}\bigg\},\hspace{-0.5 cm}
			\end{aligned}
		\end{small}
	\end{equation}
	which is a concave quadratic function of $\varphi_{j,n}$, and the optimal solution of $\varphi_{j,n}$ can be obtained by
	\begin{equation}\label{angle}
		\begin{aligned}
			\angle \varphi_{j,n}^{\text{opt}} &=\arg\hspace{-0.1 cm} \min\limits _{\theta_{n} \in[0,2 \pi)}\bigg|\theta_{n}-\angle \bigg(v_{n}-\hspace{-0.3 cm}\sum_{m \in \mathcal{N}, m \neq n}\hspace{-0.4 cm} d_{n, m} \varphi_{j,m}\bigg)\bigg| \\
			&=\angle \bigg(v_{n}-\hspace{-0.3 cm}\sum_{m \in \mathcal{N}, m \neq n}\hspace{-0.4 cm} d_{n, m} \varphi_{j,m}\bigg).
		\end{aligned}
	\end{equation}
	
	As a result, each of phase-shift can be alternately optimized by \eqref{angle} with fixing other $N - 1$ phase-shifts.
	The local optimal solution of ($\mathrm{P} 3$) can be found by iteratively updating $\boldsymbol{\varphi}_{j}$ until the convergence is achieved.
	Notice that the optimization of BS active beamforming and RIS passive beamforming is alternating in the passive beamforming design process.
	
	\subsection{User Association Optimization}
	In this subsection, we introduce the proposed dynamically successive access algorithm, which is used to optimize the user association matrix $\mathbf{U}$.
	By removing the constraints irrelevant to the user association issue, problem ($\mathrm{P} 1$) is equivalent to
	\begin{subequations}\label{P6}
		\begin{align}
			\mathrm{P} 6: \quad \mathop{\max}\limits_{\mathbf{U}} \; \quad &f_{6}(\mathbf{U})\triangleq  \sum_{k \in \mathcal{K}} R_{k},\\
			\text {s.t.} \quad
			& \eqref{uone},\eqref{oneuser},\eqref{BS},
		\end{align}
	\end{subequations}
	which is an integer programming problem.
	It can be seen in \eqref{sinr_k} that the user association variables $u_{j,k}$ cannot be extracted from the logarithm directly since the interference also depends on it.
	Inspired by \cite{J. Jiang}, we propose a dynamically successive access algorithm to optimize the user association matrix $\mathbf{U}$.
	
	To better optimize the user association, we assume that each BS serves the users with the assistance of an individual RIS, which will be designed in the dynamically access process.
	In order to simplify the objective function while maintaining a satisfactory performance, we use the SINR of user-$k$ as an equivalent metrics to determine the matching of the BS-user link. The algorithm is described in the two stages:
	
	\textit{Stage I. Benchmark user assignment:}
	
	In this stage, the main objective is to assign a benchmark user to each BS.
	To better represent the set of associated and unassociated users, we initialize the served user set of each BS as $\mathcal{A}_{j} = \varnothing, \forall j \in \mathcal{J}$ and define a set $\mathcal{Q} = \{1,...,K\}$ to denote the unassociated users.
	As the active and passive beamforming of each BS and its individual RIS have not been designed in this stage, we set the channel gain of the direct channel as the equivalent metrics.
	Then, the benchmark user-$k^{\star}$ associated with BS-$j$ can be selected by
	\begin{equation}\label{gain}
		\{j,k^{\star}\}=\operatorname{arg}\mathop{\operatorname{max}}\limits_{k \in \mathcal{Q}}  \left\|\mathbf{h}_{\text{d},j,k}\right\|^2,\forall j \in \mathcal{J}.
	\end{equation}
	
	Futhermore, when user-$k^{\star}$ is associated with BS-$j$, the set $\mathcal{Q}$, $\mathcal{A}_{j}$, and the user association matrix $\mathbf{U}$ should be updated by $\mathcal{Q}:=\mathcal{Q}\backslash k^{\star}$, $\mathcal{A}_{j}:=\mathcal{A}_{j}\cup k^{\star}$, and $u_{j,k^{\star}} = 1$, respectively.
	
	Finally, when $\left|\mathcal{A}_{j}\right|=1, \forall j \in \mathcal{J}$, the active and passive beamforming at all BS and its individual RIS should be designed by the iterative algorithm which is introduced in the previous two subsections.
	
	\textit{Stage II. Remaining user association:}
	
	In this stage, we focus on the association of the remaining users in set $\mathcal{Q}$.
	The equivalent metrics in this stage is set as the SINR.
	First, we calculate the active beamforming from BS-$j$ to the unassociated user-$k$ as $\mathbf{w}_{j, k},\forall j \in \mathcal{J}, k \in \mathcal{Q}$ by \eqref{w}, where the set $\mathcal{A}_{j}$ are replaced by $\mathcal{A}_{j} \cup k$ in this beamforming calculation process. Then, the SINR can be obtained by
	\begin{equation}\label{sinr_asso}
			\begin{aligned}
				\gamma_{j,k}=\frac{\left|\widetilde{\mathbf{h}}_{j,k}^{H} \mathbf{w}_{j, k}\right|^{2}}{\sum\limits_{i \in \mathcal{A}_j \cup k, i \neq k}\left|\widetilde{\mathbf{h}}_{j,k}^{H} \mathbf{w}_{j, i}\right|^{2}+\sigma_{k}^{2}}, \forall j\in \mathcal{J}, \forall k \in \mathcal{Q},
			\end{aligned}
	\end{equation}
	where $\widetilde{\mathbf{h}}_{j,k}^{H}$ is the equivalent channel from BS-$j$ to the remaining user-$k$.
	Based on the SINR, the BS-$j^{\star}$ and its new associated user-$k^{\star}$ can be determined by
	\begin{equation}\label{sinr_based}
		\begin{small}
			\begin{aligned}
				\{j^{\star},k^{\star}\}&=\operatorname{arg}\mathop{\operatorname{max}}\limits_{j \in \mathcal{J},k \in \mathcal{Q}}  \gamma_{j,k}.
			\end{aligned}
		\end{small}
	\end{equation}
	
	When a new BS-user pair is selected, the set $\mathcal{Q}$, $\mathcal{A}_{j^\star}$, and the user association matrix $\mathbf{U}$ should be updated by $\mathcal{Q}:=\mathcal{Q}\backslash k^{\star}$, $\mathcal{A}_{j^\star}:=\mathcal{A}_{j^\star}\cup k^{\star}$, and $u_{j^\star,k^{\star}} = 1$, respectively.
	Moreover, the active and passive beamforming of BS-$j^\star$ and its RIS should be updated.
	And the new $\mathbf{\Phi}_{j^\star}$ will be used in the next BS-user pair selection process.
	Repeat the association method introduced in stage II until $\mathcal{Q}=\varnothing$.
	
	Finally, all users will be associated with a BS after stage I and stage II.	
	
	\subsection{RIS Association Optimization}
	In this subsection, we introduce the method of optimizing RIS association vector.
	To better optimize the user association, we assume that each BS is assisted by an individual RIS in the user association algorithm. However, only one BS can be assisted by the RIS because there is only one RIS in the system and the RIS is a frequency-selective device.
	
	\begin{algorithm}[!t]
		\begin{small}
			\caption{Proposed association and beamforming design algorithm.}
			\label{alg:Algorithm 2}
			\begin{algorithmic}[1]
				\REQUIRE $\mathbf{h}_{\text{d},j,k}$, $\mathbf{G}_{j}$, and $\mathbf{h}_{\text{r}, k}$, $j \in \mathcal{J}, k \in \mathcal{K}$.
				\ENSURE   $\mathbf{U}, \mathbf{r},\mathbf{\Phi}_{j}\text{, and } \mathbf{w}_{j,k}, \forall j \in \mathcal{J}, k \in \mathcal{K}$.
				\STATE {Initialize $\mathbf{\Phi}_{j}, \forall j \in \mathcal{J}$.}
				\FOR {$j \in \mathcal{J}$}
				\STATE {Determine the benchmark user associated with BS-$j$ by \eqref{gain};}
				\STATE {Update $\mathcal{Q}$, $\mathcal{A}_j$, and $\mathbf{U}$;}
				\STATE {Initialize $\mathbf{w}_{j,k}, \forall k \in \mathcal{A}_j$ by \eqref{w};}
				\REPEAT
				\STATE {Calculate $\boldsymbol{\lambda}$ by \eqref{sinr_l} and \eqref{lamda};}
				\STATE {Calculate $\mathbf{q}$ by \eqref{q};}
				\STATE {Design $\mathbf{\Phi}_{j}$ by using the iterative reflection coefficient updating method;}
				\STATE Calculate $\mathbf{w}_{j,k}, \forall k \in \mathcal{A}_{j}$ by \eqref{w};
				\UNTIL convergence.
				\ENDFOR
				\WHILE {$\mathcal{Q}$ is not a empty set}
				\STATE {Select the new user-$k^{\star}$ associated with BS-$j^{\star}$ by \eqref{sinr_based};}
				\STATE {Update $\mathcal{Q}$, $\mathcal{A}_{j^{\star}}$, and $\mathbf{U}$;}
				\STATE {Initialize $\mathbf{w}_{j^{\star},k}, \forall k \in \mathcal{A}_{j^{\star}}$ by \eqref{w};}
				\REPEAT
				\STATE {Calculate $\boldsymbol{\lambda}$ by \eqref{sinr_l} and \eqref{lamda};}
				\STATE {Calculate $\mathbf{q}$ by \eqref{q};}
				\STATE {Design $\mathbf{\Phi}_{j^{\star}}$ by using the iterative reflection coefficient updating method;}
				\STATE Calculate $\mathbf{w}_{j^{\star},k}, \forall k \in \mathcal{A}_{j^{\star}}$ by \eqref{w};
				\UNTIL convergence.
				\ENDWHILE
				\STATE Find the optimal solutions of $\mathbf{r}$ and $\mathbf{\Phi}_{j}$ among all the exhaustion.
				\RETURN $\mathbf{U}, \mathbf{r},\mathbf{\Phi}_{j}\text{, and } \mathbf{w}_{j,k}, \forall j \in \mathcal{J}, k \in \mathcal{K}$.
			\end{algorithmic}
		\end{small}
	\end{algorithm}

	When the solution of $\mathbf{U}$ is obtained by the dynamically successive access algorithm, all the optimal $\mathbf{\Phi}_{j}$ for each BS have been obtained.
	Then, all the possibilities of $\mathbf{r}$ that satisfies constraint \eqref{rone} and \eqref{RIS} can be examined. Based on the optimal $\mathbf{\Phi}_{j}$, the sum-rate of each possibility can be obtained.
	Then, the optimal solution of $\mathbf{r}$ can be selected.
	With this exhaustive search method, the complex problem caused by the coupling in numerator and denominator between $\mathbf{r}$ and $\mathbf{\Phi}_{j}$ can be easily solved.
	
	With the above described solution for each sub-problem, we can solve the original problem ($\mathrm{P} 1$) by the proposed algorithm, which is summarized in Algorithm \ref{alg:Algorithm 2}.	
	
	\section{Simulation Results}
	In this section, the simulation results are provided to evaluate the performance of the proposed BS-RIS-user association and beamforming design algorithm for the RIS-aided cellular networks.
	We consider that $J = 4$ BSs are located at $\left(0 \text{ m},200\text{ m} \right) $, $\left(-150 \text{ m}, 0\text{ m} \right)$, $\left(250 \text{ m},0\text{ m} \right)$, $\left(0 \text{ m},-300\text{ m} \right)$, respectively.
	The RIS is deployed at $\left(0 \text{ m}, 0\text{ m} \right)$ and $K = 25$ single-antenna users are randomly distributed in a circle centered at $\left(25 \text{ m},-25\text{ m} \right)$ with radius $150$ m.
	Each BS is equipped with $M = 32$ antennas and the RIS consists of $N = 64$ reflection elements.
	The noise power $\sigma_{k}^2$ at the receivers and the maximum tranmit power $P_{\text{max}}$ are set to $-80\text{ dBm}$ and $50\text{ dBm}$, respectively.
	
	We assume that the BS-user channels are the Rayleigh fading channels and the BS-RIS, RIS-user channels are Rician channels.
	Since the RIS is deployed on high buildings, there are strong line of sight (LoS) link between BSs and RIS.
	Similarly, the RIS-user link will also contain the LoS link. Thus, we set the rician factors as $\infty$ and 1 for the BS-RIS and RIS-user channels, respectively.
	The path loss depends on the distance and can be calculated by
	\begin{equation}
		\begin{small}	
			L(d)=C_{0}\left(\frac{d}{D_{0}}\right)^{-\alpha},
		\end{small}	
	\end{equation}
	where $C_{0}$ and ${D_{0}}$ are -30 dB and 1 m, respectively \cite{Q. Wu}.
	The path loss exponent $\alpha$ is set to 3.9, 2.5 and 2.7 for the BS-user, BS-RIS, and RIS-user channels, respectively.
	
	\begin{figure}[!t]\vspace{-0.4cm}
		\centering
		\includegraphics[width=3.1 in]{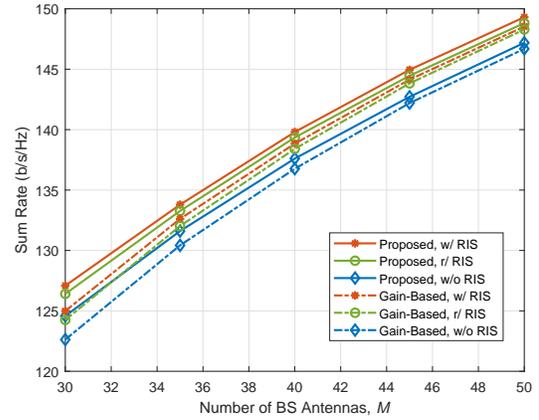}
		\vspace{-0.15 cm}
		\caption{Sum-rate versus the number of BS antennas $M$.}
		\vspace{-0.5 cm}
		\label{fig:M}
	\end{figure}
	In order to better verify the effectiveness of the proposed algorithm and ensure the fairness of the comparison, we compare the proposed algorithm with the direct channel gain based (Gain-Based) user association algorithm \cite{P. Ni} in the following three situations:
	
	i) With RIS (w/ RIS): A RIS is deployed and the passive beamforming is designed based on the proposed algorithm;
	
	ii) Random RIS (r/ RIS): A RIS is deployed and the passive beamforming is randomly generated;
	
	iii) Without RIS (w/o RIS): No RIS is deployed.
	
	Fig. \ref{fig:M} shows the sum-rate of all users as a function of the number of BS antennas $M$.
	It can be observed that with the increasing of $M$, the sum-rate also increases.
	Meanwhile, the proposed algorithm can provide better performance than the other algorithms.
	Although the performance of Gain-Based algorithm improves with the increasing of $M$, it is still worse than the proposed algorithm.
	\begin{figure}[!t]\vspace{-0.4cm}
		\centering
		\includegraphics[width=3.1 in]{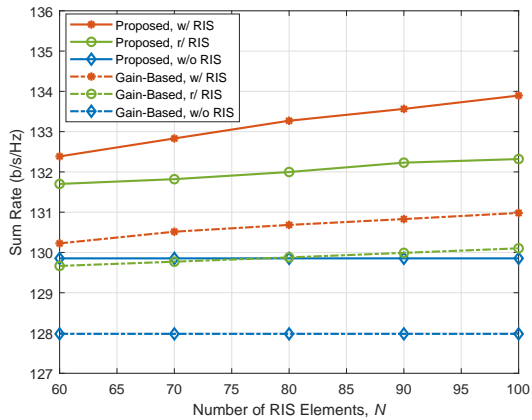}
		\vspace{-0.15 cm}
		\caption{Sum-rate versus the number of RIS elements $N$.}
		\vspace{-0.5 cm}
		\label{fig:N}
	\end{figure}
	\begin{figure}[!t]
		\centering
		\includegraphics[width=3.1 in]{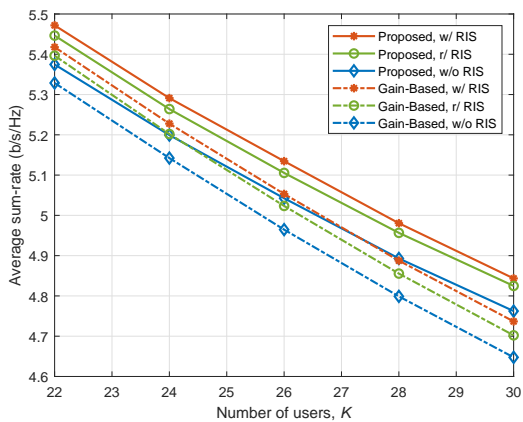}
		\vspace{-0.15 cm}
		\caption{Average sum-rate versus the number of users $K$.}
		\vspace{-0.5 cm}
		\label{fig:K}
	\end{figure}
	
	Fig. \ref{fig:N} illustrates the relationship between sum-rate and the number of RIS reflection elements $N$.
	It is obvious that the sum-rate increases with the increase of RIS passive elements $N$.
	Moreover, the proposed algorithm can achieve better performance among the other methods.
	Since the main difference is the user association algorithm, it is easy to conclude that the dynamically successive access algorithm is more efficient.
	
	Fig. \ref{fig:K} describes the average sum-rate (i.e., $\sum_{k \in \mathcal{K}}R_k/{K}$) performance with respect to the number of users $K$.
	When there are more users in the system, the average sum-rate will decrease because of the increasing co-channel interference and the decreasing power which is allocated to each user.
	With the increasing number of users, the performance superiority of the proposed algorithm is more prominent.
	Meanwhile, the performance of Gain-Based algorithm is even worse than the w/o RIS situation which dynamically optimizes BS-user links.
	
	\section{Conclusion}
	This paper aims to maximize the sum-rate of all users in a RIS-assisted cellular system by jointly optimizing the BS-RIS-user association as well as the active and passive beamforming.
	To solve the non-convex problem, we propose a dynamically successive access user association algorithm, an iterative algorithm, and an exhaustive search method.
	The effectiveness of the proposed algorithm is verified in the simulation results.
	Moreover, with the increasing variables, the proposed algorithm can attain better sum-rate performance than the Gain-Based algorithm, especially when the number of users increases.

\end{document}